\documentstyle[prl,aps,twocolumn]{revtex}

\begin{document}
\title{Universal Quantum Entanglement Concentration Gate}
\author{Zi-Yang Wang, Chuan-Wei Zhang, Chuan-Feng Li, Guang-Can Guo\thanks{%
Electronic address: gcguo@ustc.edu.cn}}
\address{Laboratory of Quantum Communication and Quantum\\
Computation and Department of Physics,\\
University of Science and Technology of China,\\
Hefei 230026, People's Republic of China\vspace*{0.3in}}
\maketitle

\begin{abstract}
\baselineskip16ptWe construct a Universal Quantum Entanglement Concentration
Gate (QEC-Gate). Special times operations of QEC-Gate can transform a pure
2-level bipartite entangled state to nearly maximum entanglement. The
transformation can attain any required fidelity with optimal probability by
adjusting {\it concentration step}. We also generate QEC-Gate to the Schmidt
decomposable multi-partite system.

PACS numbers: 03.67.-a, 03.65.Bz, 89.70.+c \\ 
\end{abstract}

\baselineskip16pt The entanglement nature is a key source to distinguish
quantum and classical information theory. Some authors $[1-3]$ have been
successful in identifying many fundamental properties of entanglement.
Because many important quantum processes, such as teleportation $[4]$,
superdense code $[5]$ and quantum computation $[6]$, require maximally
entangled states, the methods of entanglement enhancement and entanglement
concentration was frequently discussed $[1-3,7-9]$. Such work is closely
related to the fundamental problems of state transformation $[10-18]$. Two
main protocols, probabilistic $[7,11]$ and approximate $[16]$, has been
applied in the optimal process. An unavoidable problem of these work is that
the transformation greatly depends on the input states.

In this report, we construct a Universal Quantum Entanglement Concentration
Gate (QEC-Gate), multi-operation of which can transform a $2$-level
bipartite partial pure entangled state to a nearly maximally entangled state
with optimal probability. Its fidelity can be entirely controlled by the
parameter {\it concentration step}. Any required fidelity can be approached
by adjusting {\it concentration step} and enhancing operation times.
Additionally, each operation of QEC-Gate can always process with the optimal
probability.

In general, a $2$-level bipartite entangled state can be represented as the
following according to Schmidt decomposition, 
\begin{equation}
\left| \psi \right\rangle =\cos \theta \left| 11\right\rangle +\sin \theta
\left| 00\right\rangle ,  \eqnum{1}
\end{equation}
where $0\leq \theta \leq \frac \pi 4$. Its entanglement, $E\left( \left|
\psi \right\rangle \right) =-\cos ^2\theta \log _2\cos ^2\theta -\sin
^2\theta \log _2\sin ^2\theta $ can be compared directly by $\theta $.

Suppose the pure entangled state is shared by two distant observers, Alice
and Bob. Since any operation in quantum mechanics can be represented by a
unitary evolution together with a measurement, a probe $P$ with a Hilbert
space, $\left| P_0\right\rangle \otimes \left| P_1\right\rangle $ ($\left|
P_0\right\rangle $ and $\left| P_1\right\rangle $ are orthogonal), is
introduced. Our gate operation is a local unitary evolution (Alice side)
together with the postselection of the measurement results which can be
generally represented as
\begin{equation}
(\hat U_{AP}\otimes I_B)\left| \psi \right\rangle \left| P_0\right\rangle =%
\sqrt{\gamma }\left| \tilde \psi \right\rangle \left| P_0\right\rangle +%
\sqrt{1-\gamma }\left| \phi \right\rangle _{AB}\left| P_1\right\rangle , 
\eqnum{2}
\end{equation}
where $\gamma $ is the probability of successful state transformation of $%
\left| \psi \right\rangle \rightarrow \left| \tilde \psi \right\rangle $, $%
\left| \phi \right\rangle _{AB}$ is an arbitrary state in the composite
system of Alice and Bob. If the measurement of the probe results in $\left|
P_0\right\rangle $, this transformation is successful.

The QEC-Gate $G\left( \xi ,\eta \right) $ executes the unitary-reduction
operation $U_{AP}$ which is described as the following with two parameters $%
\xi $ and $\eta $, $0<\xi <\eta <\frac \pi 4$,
\begin{eqnarray}
\hat U_{AP}\left| 1\right\rangle _A\left| P_0\right\rangle  &=&\sqrt{\gamma
_0}\frac{\cos \eta }{\cos \xi }\left| 1\right\rangle _A\left|
P_0\right\rangle   \eqnum{3} \\
&&+\sqrt{1-\gamma _0}\frac 1{\cos \xi }\left| 1\right\rangle _A\left|
P_1\right\rangle ,  \nonumber
\end{eqnarray}
\begin{equation}
\hat U_{AP}\left| 0\right\rangle _A\left| P_0\right\rangle =\sqrt{\gamma _0}%
\frac{\sin \eta }{\sin \xi }\left| 0\right\rangle _A\left| P_0\right\rangle ,
\eqnum{4}
\end{equation}
where $\gamma _0=\frac{\sin ^2\xi }{\sin ^2\eta }$ is exactly the optimal
probability to transform entangled state $\cos \xi \left| 11\right\rangle
_{AB}+\sin \xi \left| 00\right\rangle _{AB}$ to $\cos \eta \left|
11\right\rangle _{AB}+\sin \eta \left| 00\right\rangle _{AB}$ $[11]$. In
fact, $U_{AP}$ can be implemented by qubit $A$ controlling $P$ rotation. It
can be expressed as a generalized Toffoli gate $\Lambda _1\left( U\right) $ $%
\left[ 19\right] $ that applies $U$ to qubit $P$ if and only if $A$ on $%
\left| 1\right\rangle $, where $U=\left( 
\begin{array}{cc}
\delta  & -\sqrt{1-\delta ^2} \\ 
\sqrt{1-\delta ^2} & \delta 
\end{array}
\right) $ with $\delta =\frac{\tan \xi }{\tan \eta }$.

We set $\theta =\theta _0\neq 0,\frac \pi 4$ in Eq.(1) as the partial
entangled initial input state (it is reasonable because if the initial input
state's Schmidt decomposition is represented on another base, an unitary
operation $V_A\otimes V_B$ is needed to transform it to the form of Eq.(1)).
One operation of $G\left( \xi ,\eta \right) $ on the input can be
represented as
\begin{eqnarray}
&&\left( \hat U_{AP}\otimes I_B\right) \left( \cos \theta _0\left|
11\right\rangle _{AB}+\sin \theta _0\left| 00\right\rangle _{AB}\right)
\left| P_0\right\rangle   \eqnum{5} \\
&=&\sqrt{\gamma _0}\left( \cos \theta _0\frac{\cos \eta }{\cos \xi }\left|
11\right\rangle _{AB}+\sin \theta _0\frac{\sin \eta }{\sin \xi }\left|
00\right\rangle _{AB}\right) \left| P_0\right\rangle   \nonumber \\
&&+\sqrt{1-\gamma _0}\frac{\cos \theta _0}{\cos \xi }\left| 11\right\rangle
_{AB}\left| P_1\right\rangle .  \nonumber
\end{eqnarray}

The normalization of output state in Eq.(5) can yield the transformation
probability $\gamma _1=\frac{\sin ^2\theta _0}{\sin ^2\theta _1}$. This
probability is also optimal according to Vidal Theorem $\left[ 11\right] $.

Denote the output entangled state as $\cos \theta _1\left| 11\right\rangle
_{AB}+\sin \theta _1\left| 00\right\rangle _{AB}$, we get 
\begin{equation}
\tan \theta _1=\frac 1\delta \tan \theta _0.  \eqnum{6}
\end{equation}

Define {\it Concentration Step }$\Delta (\xi ,\eta )=\frac 1\delta =\frac{%
\tan \eta }{\tan \xi }$. Notice that if $\xi $ and $\eta $ are set properly,
which means that $\Delta \left( \xi ,\eta \right) -1$ has a small value, we
can promise the enhancement of entanglement. A natural application is to
operate $G\left( \xi ,\eta \right) $ on the initial input state more than
one times. For each operation, the entanglement get an increase. Surely this
situation can not continue infinitely and the converting operation indicates
the approach of maximum. Another obvious property of QEC-Gate is that the
enhancement of entanglement is discrete. So in general cases, the final
concentrated state is not exactly the maximum entangled state. Here we can
determine the optimal operation times $T$, the final output state $\left|
\Phi _{final}\right\rangle $, total probability $\Gamma $ and fidelity $F$.

\begin{equation}
T=\left[ -\frac{\ln \tan \theta _0}{\ln \Delta \left( \xi ,\eta \right) }%
\right] ,  \eqnum{7}
\end{equation}
where $\left[ x\right] $ is the Gauss Functor which gives the integer part
of real number $x$. Notice that this result is based on the limit of $\theta
_T\in \left( 0,\frac \pi 4\right] $. If we permit $\theta _T>\frac \pi 4$,
when $\theta _i\leq \frac \pi 4<\theta _{i+1}$, we set $T=i$ if $\left|
\theta _i-\frac \pi 4\right| \leq \left| \theta _{i+1}-\frac \pi 4\right| $,
otherwise, $T=i+1$. This choice can promise the optimal output.
\begin{eqnarray}
&&\left| \Phi _{final}\right\rangle   \eqnum{8} \\
&=&G^T\left( \cos \theta _0\left| 11\right\rangle _{AB}+\sin \theta _0\left|
00\right\rangle _{AB}\right)   \nonumber \\
&=&A\left( \cos \theta _0\left( \frac{\cos \eta }{\cos \xi }\right) ^T\left|
11\right\rangle _{AB}+\sin \theta _0\left( \frac{\sin \eta }{\sin \xi }%
\right) ^T\left| 00\right\rangle _{AB}\right) ,  \nonumber
\end{eqnarray}
where $A$ is a normalizing coefficient which is 
\begin{equation}
A\left( \theta _0\right) =\left( \left( \cos ^2\theta _0+\Delta ^{2T}\sin
^2\theta _0\right) \left( \frac{1+\tan ^2\xi }{1+\tan ^2\eta }\right)
^T\right) ^{-\frac 12}.  \eqnum{9}
\end{equation}

The probability and fidelity are 
\begin{equation}
\Gamma =\prod \gamma _i=\frac{\sin ^2\theta _0}{\sin ^2\theta _T}\rightarrow
2\sin ^2\theta _0,  \eqnum{10}
\end{equation}
\begin{equation}
F=\frac 12\left( \frac{1+\tan \theta _0\Delta ^T\left( \xi ,\eta \right) }{%
\sqrt{1+\tan ^2\theta _0\Delta ^{2T}\left( \xi ,\eta \right) }}\right) ^2. 
\eqnum{11}
\end{equation}

From the results above, we can see that the total probability is only
related to the initial input state. The probability of each transformation
approaches $1$ when $\Delta \left( \xi ,\eta \right) -1$ is small enough.
But with the increasing operation times, $\Gamma $ remains approximately
unchanged.

Because operation times $T$ is not a continuous variable, the fidelity
represented by Eq.(10) changes periodically with $\theta _0$. The period is
determined by {\it concentration step} $\Delta $, that is $F\left( \theta
_0\right) =F\left( \theta _0\arctan \Delta \right) $. For some special $%
\theta _0\in \left\{ \theta _0^{\left( k\right) }=\arctan \frac 1{\Delta ^k}%
,k=1,2,...\right\} $, $F$ can exactly attain 1. But the average fidelity, or
even the minimum, is more important. To examine the relationship of $F$ and 
{\it concentration step}, we firstly calculate the minimum of $F\left(
\theta _0\right) $. In this situation, $\frac \pi 4-\theta _T=\theta _{T+1}-%
\frac \pi 4$, so we get 
\begin{equation}
F_{\min }\left( \Delta \right) =\min F\left( \theta _0\right) =\frac 12%
\left( \frac{1+\sqrt{\Delta }}{\sqrt{1+\Delta }}\right) ^2.  \eqnum{12}
\end{equation}

The differential coefficient shows that $F_{\min }\left( \Delta \right) $ is
a monotonous decreasing function of $\Delta $. We can increase the fidelity
by reducing $\Delta \left( \xi ,\eta \right) $. When $\Delta \rightarrow 1$, 
$F_{\min }\left( \Delta \right) \rightarrow 1$. Obviously, it is realized at
the price of increasing operation times.

We (three of us) $[18]$ have shown that two different entangled states can
be concentrated by same local operations and classical communication if and
only if they share the same marginal density operator on one side. Although
the design of QEC-Gate makes it possible to attain any required fidelity, it
doesn't contradict with the previous result. For different initial states,
the final states and the operation times, which determine the practical
operation of QEC-Machine, may be different. Actually, the discrete strategy
which is adopted to simulate the continuous transformation space can create
an identical unit to approximately measure the space. The finer the
separation is, the better the measure is. This is exactly the reason why a
universal gate can be constructed.

For the complicated situation of $n$-partite entanglement, the QEC-Gate can
not be directly generated. But it is still efficient in transforming some
special entanglements, such as the Schmidt decomposable entangled states $%
\left\{ \cos \theta \left| 11...1\right\rangle +\sin \theta \left|
00...0\right\rangle ,0<\theta <\frac \pi 4\right\} $. Since the QEC-Gate
operates only on the composite system of one of the $n$ parties and the
probe, the result is the same as that of $2$-partite state transformation.

In summery, we have constructed a Universal Quantum Entanglement
Concentration Gate, multi-operation of which can transform a partial
entanglement to a nearly maximum entanglement with optimal probability. Any
required fidelity can be approached by decreasing {\it concentration step}.

This work was supported by National Nature Science Foundation of China.

\end{document}